\documentclass[twoside]{article}

\usepackage{epsf}
\usepackage{latexsym}
\usepackage{times}
\usepackage{amssymb}
\usepackage{csit}

\title{\bf L.T.Kuzin: Research Program}
\author{\large Viacheslav Wolfengagen\\[1.52mm]
\normalsize Institute for Contemporary Education ``JurInfoR-MSU'',\\
\normalsize Vorotnikovsky lane, 7, bld. 4\\
\normalsize Moscow, 103006 Russia
\normalsize {vew@jmsuice.msk.ru}\date{}}

\institution{}
\begin{document}
 \setcounter{page}{97}

\markboth{L.T.Kuzin: Research Program}
{Workshop on Computer Science and Information Technologies CSIT'99,
Moscow, Russia, 1999}

\maketitle

\epsfxsize=3.2in
\centerline{\epsfbox{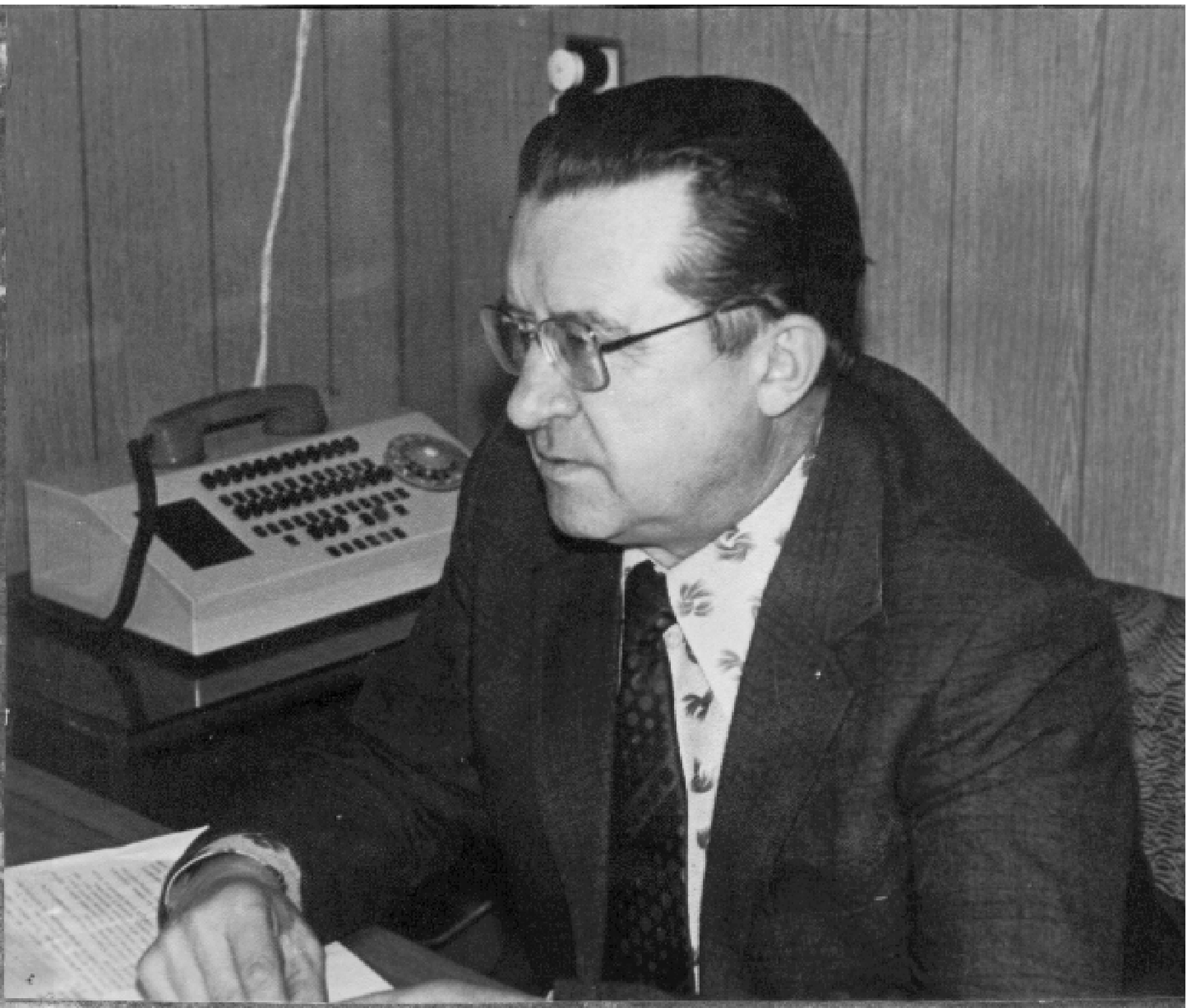}}
\medskip
\centerline{Dedicated to the memory of Professor Lev Kuzin}
\centerline{on the occasion of his 70th Birthday}

\begin{abstract}
\noindent Lev T. Kuzin (1928--1997) is one of the founders of
modern cybernetics and
information science in Russia.
After getting his Doctor of Technical Sciences
degree in the industrial research sphere he joined the Moscow Engineering
Physical Institute at the time when there were a few specialists in
cybernetics in Russia. A few years later he became a founder of Cybernetics department
and a Chair of the section in Applied Problems for Cybernetics in Moscow House
for Scientific and Technical Knowledge Propagation. Almost the same time
he became a project principal researcher
(project {\em Data Bank} promoted by the Russian Academy of Sciences),
and attracted the teams of researchers to apply the computational models to a field
of data and knowledge bases, making the pioneering research in Artificial Intelligence.

He was awarded and honored the USSR State Prize
for inspiring vision into the future of technical cybernetics
and his invention and innovation of key technologies.

The last years he interested in the computational models
of geometrical and algebraic nature
and their applications in various branches of computer science
and information technologies.
In the recent years the interest in computation models
based on object notion has grown tremendously stimulating an interest
to Kuzin's ideas. This year of 50th Anniversary of Cybernetics
and on the occasion of his 70th birthday on September 12, 1998
seems especially appropriate for discussing Kuzin's Research Program.
\end{abstract}

\section{Introduction}

The explanation to the term {\em research program of L.Kuzin}
arose from the attempts to analyze the trends, directions
and stages of his scientific activity.
Kuzin did not used this expression in its own sense but often
preferred some way of making the research and development work.

The obvious and simplest way is to declare that the general
topic of his research was cybernetics and information science.
Nevertheless, this explanation is not precise and exact.
For better acceptance and understanding of his efforts,
interests and activity
it is important to pay attention to his basic aims and approaches.
In case we forgot this general direction we have a chance to loose
this understanding.

There has been some general way to understand and select out
his research road. But there has been misunderstanding of his
main aims and general research approach.
One of the reasons for this is that many of his, especially early,
tutorials, invited talks and keynote papers appeared
in a few copies published in
widely inaccessible paper collections or
even with a limited access to copies. His ideas have changed
with a time but the later ideas, as a rule, arose from
the earlier preliminary contributions and projects.
The evolution of his ideas with a convenience can be
divided into four periods: 1)~till early 60s,
2)~from 1963 till 1975, 3)~from 1975-1976 till 1983,
and 4) after 1983-1986.

 \section{Period till early 60s}
  \label{chapt:01}
Kuzin earned M.S. in Radio-physics as an Engineering Mathematician from Gorky State
University in 1951. For 15 years he worked
as Senior Scientist and Principal Scientist
at the military research
institutes where he earned his Ph.D. degree and Doctor of Technical
Sciences degree.

This ten years period he worked in military research institutions
at the governmental tasks in a field of technical cybernetics.
He characterized his position as the
``fitter and faultfinder of 10 thousands electronic tubes''.
He acquired a good experience in developing, fitting and inculcating of
the large complex technical systems. But he felt a discomfort with a gap
between the theory and engineering practice. On the one hand a theory was extremely abstract,
long distant from the practical needs. On the other hand the developers of the complex systems
had no effective engineering methods because of their lack. In particular, the educational
background of the specialists was insufficient for a success of the
development, they had an extreme
needs for good background knowledge
in system analysis, operation research, and cybernetics.
That is why the research and development efforts were often directed to a blind alley,
when this became obvious they were canceled out and re-initiated from the starting point.
During five years he worked as a principal investigator in a theory of large complex technical
systems, and he observed a frequent change of the senior managers for their failed projects.

All of this inspired his decision to change his position,  join an educational institute and start the
academic activity to contribute to the educational programs for managers, developers of the
complex technical systems with well balanced engineering and theoretical ground. This time he
was invited to join the Moscow Engineering Physical Institute and found a new department.
In establishing the educational schedules he used his experience from Gorky State University,
where due to the school of academician A. Andronov, his Professor,
in 1946 the first in USSR
technical radio-physics faculty aimed to educate the
engineering mathematicians was established.
A feature of this educational direction was the accommodation of existing methods and
developing the new theoretical approaches to solving the real tasks
arising during the
development of a large complex technical system. The educational system has a lack of the clear
written source books in cybernetics which were written at a good engineering level without a lost
of mathematical generality, and Kuzin decided to start this new stage of his career.

 \section{Period from 1963 to 1975}
   \label{chapt:02}
Lev Kuzin was awarded the USSR State Prize. Kuzin was honored
for inspiring vision into the future of technical cybernetics
and his invention and innovation of key technologies.

For nearly 35 years he worked permanently
at Moscow Engineering Physical Institute
where he founded the department initially named
``Managing computing machines''
and later renamed as ``Cybernetics''.

During the first two years he concentrated his efforts at the detailed preparations of the academic
programs, plans and schedules in the new speciality to launch the long term activity for educating
the engineering mathematicians.

He and his collaborators  worked closely with industry, academia
and governmental structures to promote a principal improvement
of information infrastructure strongly oriented towards
processing of the very large information bulks and
later very large data bases and knowledge processing.

This was a stage in his career when he interested in interactive
computing. He observed that in 60s and 70s automatic computing
in Russia was mainly valued for data processing by the industrial
communities.
In 70s his interests moved permanently into the computer field,
he paid a lot of attention of humans using computers interactively
via CRT displays. After that he had become convinced that
computers could make a difference between the humans way of
thinking, everyday engineering reasoning and computational plans,
so that he decided to contribute in handling the volumes
of professional knowledge. He felt that this direction
of research has a long range future and perspective
and a lot of impacts towards a higher education.

 \section{Period of 1975--1983}
   \label{chapt:03}
In the 1973--1976 period he wrote a series of reports,
in which he set out a conceptual
framework about how to organize the medium and environment where
the humans could actually represent their knowledge.
Later he was under an
influence of ARPA whose leaders committed to the idea of
establishing interactive computing with time-sharing.

By 1980 he was one of the principal investigators with a series
of industrial projects and started working on a project on
`intelligent computing'. This time he was deeply interested
in trying to recruit a vast community of people to start learning
how to do the collaborative and multistage investigation of
knowledge discovering and representation. He had some good teams
of volunteers who started the partial research in a field. Often they
worked days-and-nights to concentrate on fastest achievement
even of the intermediate results and working software prototypes.

One of his proposals in 80s was to develop an extensible
knowledge representation environment. His advice to the research
community was to implement an enhanced OPS Information Technology,
and a promising project was implemented by his collaborator
Oleg Balovnev. This was the full scale OPS computational environment,
the toolkit, information technology and
corresponding particular applications were used for knowledge
systems at the industrial enterprises giving rise to
a fruitful research stream.

Nevertheless, in the first half of 70th most of the computing ideas
moved more in the direction how to automate the industrial routine work
with the huge databases than how to represent and discover the knowledge
and establish the knowledge base which would be able to communicate
with the databases. This period he launched several Ph.D. research topics
to cover the problems of long-range future. He kept thinking about
the humans with a high and deep professional training whose `natural'
knowledge could be transformed with the aid of knowledge engineers
into a suitable computing schemata. He thought both of declarative and procedural
approach trying to combine them into the uniform computational environment.

 \section{After 1983}
   \label{chapt:04}
He felt that the professional in all fields could boost the quality of their knowledge
work using the knowledge representation engines. But for a long time
it was the single focus of his efforts, because there were a lot of years
when he kept trying to save and continue the most promising working
projects which were already launched and running in industry.

During the time in the 70s till the mid-80s he accumulated
his experience, knowledge and intuition
in a field of information systems in general. He observed the most important
projects in government, academy, various research fields, industry to understand
the scale and rate of their changes with a time.
In each of the branches the changes were with external/internal knowledge schemata
and operations, the challenge was for interoperable information exchange
using, possibly, the relational databases and meta-relational knowledge bases.
He was influenced by the 5th Generation Computing Systems project and tried to
select out the basic, but outstanding and challenging ideas.

He used his Ph.D. students research as a platform from which,
using the prototype software systems, to launch the promising computational
ideas into industry. He felt importance of collaborative efforts with different
participating organizations to cooperate in order to explore the possible
and actual ways in which the technical systems of artificial intelligence
can be done.

Intermediately he and his collaborators ended up generating a concept
that they called the ``Intelligent Data Bank'', or shortly, IDB.
He described the real phenomenon
of transactions between the counterparts of this poli-based information
human-machine system. This time Kuzin and his team launched a series
of conferences and symposia to broader discuss  the aims, targets and results in
this direction. His main idea was based on the effects of group of experts
behavior when they act collaboratively and try to get familiar with
their environment and to coordinate the common resources with
a purposeful efforts and actions. Under these circumstances they
must integrate their fuzzy and incomplete knowledge into the entire IDB.

From the mid-80s with a group of supporters he kept pitching the topics
of the intermediate direction between the computer science and
information technologies.
In particular, he was interested in an idea to apply the
categorical universe of discourse to
the field of data models which erased in the late 70th -- early 80th.

He paid special attention
to discuss applications of categories to the field of data
modeling\footnote{See, for example, his books: L.T.Kuzin,
{\em Foundations of Cybernetics}, Vol. 2, Moscow, Energija, 1979,
and L.T.Kuzin, {\em Foundations of Cybernetics},
Vol. 1, Moscow, Energija, 1994, both printed in Russian.}.
His early idea was to use some kind of untyped calculi and impose some
weak restrictions to capture the intutitive reasons concerning
the data models. This approach tends to taking in mind a family
of {\em computation models} as the restrictions of the untyped
combinatory logic or untyped $\lambda$-calculus.

During more than a decade he worked as a project principal researcher
(project {\em Data Bank} promoted by the Russian Academy of Sciences
during 70th -- early 80th), he attracted the teams of researchers
to apply the computational models to a field of data and
knowledge bases.
Having launched in Russia some research projects in a field he discovered
that the vital point of a general data model is to build an access
to the computation environment. His proposal was to separate the
{\em function} symbol and an {\em argument} symbol
(when the functional expression is included within a query),
then to generate access to the domain for argument and to
the domain for function separately. The most of the dufficulties
at this way were observed and some of them were resolved.

Later the more neutral approach was established, roughly speaking,
to deal with all the {\em both-syntax-and-semantic} entities
as with the homogeneous
{\em objects}. The objects are interlinked by the relations which
correspond to the {\em scripts} (script-driven data model).
This model was extended to capture more {\em dynamics}. Studying
the transitions between the scripts he verified some pure
categorical models (with the explicit notion of a {\em state}
in a computation model).

Nevertheless, the {\em semantics} of the computation in a category
was not yet covered. During the last period of his research
 he was interested in discovering more and more
facts which  fit in the mathematical ideas beyond a category theory.
In his draft research schedule he had done some improvements
concerning the feasibility of a categorical abstract machine
to evaluate the queries.

 \section{Selected publications of Lev T.Kuzin}
   \label{chapt:05}
This is not a complete list of Kuzin's contributions, reports, papers,
and books. A lot of his reports were published just as the very short
theses in hardly available and accessed collections. A part of them
were known in details as manuscripts. Another part was represented
only as a collections of transparencies which he used during his
regular reports at the monthly seminar ``Artificial Intelligence''
in Moscow House for Scientific and Technical Knowledge Propagation
which existed more than a decade. Many of his papers were
contributed with the collaborating co-authors. This list is compiled
to reflect Kuzin's most promising ideas and challenges,
and corresponds to the list of the event which he inspired or stimulated.

In addition, Kuzin was Editor-in-Chief of the periodical
``Engineering matematical methods for physics and cybernetics''
(Atomizdat, Moscow, Russia).
His aims were to stimulate, step by step, the research activity
and publications from an open position
of Information Technologies. The main target
was to launch the development of engineering mathematical
methods based on the computer technologies in the various
scientific fields. He applied efforts to select out
and review the contributions in a field of cybernetics.
This periodical has attracted the research communities
and stimulated their activity in a manner
and in an area of engineering mathematicians.

 \section{Some of the prominent events stimulated,
initiated and/or inspired by Lev Kuzin}
   \label{chapt:06}
This list includes some of the most important events which were inspired,
stimulated and/or initiated by Lev T. Kuzin's and is not complete.
The comments to his publications and contributions are shortly given
because some of his works were  never published as a complete papers,
or were published in a minimal amount of copies. His most important
publications are printed in Russian.

\vspace{1ex}
\centerline{\bf 1963}
\vspace{1ex}

\noindent He founded at Moscow Engineering Physical Institute
the department No 22 ``Controlling electronic computing machines''
later renamed as ``Cybernetics''. A few time later he was awarded
the USSR State Prize for contributing in technical cybernetics.

\vspace{1ex}
\centerline{\bf 1971}
\vspace{1ex}

\noindent {\sc Semiotics methods for managing  the large systems.}
Seminar at the Moscow House for Scientific and Technical
Knowledge Propagation (MHSTKP)

\vspace{1ex}
\centerline{\bf 1973}
\vspace{1ex}

\noindent He analyzed the state of implementing and applications of
Management Information Systems (MISs) in the USSR
at the seminar with government representatives. He gave a system analysis
of MIS in a scale of the big city like Moscow.
This was aimed to effective using of computers and mathematical methods
in industry and research institutions.

\vspace{1ex}
\centerline{\bf 1974}
\vspace{1ex}

\noindent {\sc Artificial Intelligence. Advances and Perspectives}.
Seminar at the Moscow House for Scientific and Technical
Knowledge Propagation (MHSTKP)

\vspace{1ex}
\centerline{\bf 1976}
\vspace{1ex}

\noindent He was one of the principal officers at the USSR conference
{\sc Management Information Systems} on May 10-13, Tbilisi, Georgia.

This was a forum for the scientists of the country.
He had the invited talk on the problems of modeling for MIS.
He summed up the results of applying the continuous mathematics
under the new conditions in developing MISs. Kuzin paid attention
to a strong necessity to use and establish the mathematical models
of other types which previously were unknown because
``\dots {\em the actual and existing industrial conditions need the
frequent changes of criteria,
the ranges of possible values for controlled parameters, and
the functional structure and restrictions are changeable.
The usual means from the general mathematical ware for computer
do not give to the specialist from industry
an ability to find out the controls which are optimal under the ad hoc
industrial situation, without troublesome  interference of the mathematicians
and programmers}''.

One of the possible solutions on his opinion were DBMSs.
In particular, he had a serious interest to the relational databases,
he inspired and personally assisted the research in this area.
Usually he gave to the beginner in this field all the needed support
helping to overcome both the principle scientific and purely practical difficulties.

At this event he moderated some round table discussions.
Many of his listeners later became the serious researchers and high
skilled specialist both in MIS software and in relational DBMS
development and applications.


\vspace{1ex}
\centerline{\bf 1977}
\vspace{1ex}

\noindent {\sc The 1-st USSR workshop on Intelligent Data Banks}.
He was a General Chair at the 1-st USSR workshop
``Intelligent Data Banks'', Sukhumi, Abkhazia.

In his tutorial Kuzin started up generating a concept of the
Intelligent Data Bank (IDB). He determined this kind
of information systems as ``\dots {\em having an ability to generate
or derive the new information which was not previously
explicitly present, and having the following properties:
interface via professionally restricted natural language,
multi-user mode, self-learning via open dialog interface}''.

In fact, this was a premier workshop in USSR in this direction
which was aimed to disseminate the AI ideas all over the country.
In particular,  Professors G.G.Chogowadze, G.G.Gogichaishwili
re-directed their collaborators in Tbilisi, Georgia, towards
more close usage of the AI principles and approaches.
They were among the founders of the Georgian national school of Artificial
Intelligence.

\vspace{1ex}
\centerline{\bf 1978a}
\vspace{1ex}

\noindent {\sc Computer Aided Design}.
He was a moderator and one of the General Chairs at the Moscow seminar
``Computer Aided Design'' at the Moscow House for Scientific and Technical
Knowledge Propagation (MHSTKP).

Kuzin inspired this large scale seminar where many of the known specialists
and the beginners in a field took part.
A dominant trend was to discuss the results of finished and not-finished
research projects and difficulties with the implementations and applications.
Among them was a series of projects which were implemented by
Kuzin's collaborators,   this was an actual outcome of Kuzin's school.

In his tutorial he discussed a conceptual framework for the CAD system
with the intelligent behavior, interface and other properties. This time he
actively worked in this area and was a principal researcher in this branch.
He sought for the opportunities to integrate the implementations
of various logic means with the methods. This time he started to formulate
the software design principles
in terms of objects and scripts, he initiated and
started the implementation of a tool kit based on event-driven doctrine.

He understood an importance to develop the general technology and discipline
of programming, usage of the logical filters which would give more freedom
to a usual or causal user. Most of his ideas were accepted by the USSR's specialists,
they were used very naturally and  in the most of the research and industrial
communities.

This conceptual transparency of his tutorial, extremely natural in its origin,
caused the wide dissemination,  within the whole country,
of the approaches and methods discovered by the collaborative
groups of the researches, and first of all by Kuzin's school.

\vspace{1ex}
\centerline{\bf 1978b}
\vspace{1ex}

\noindent {\sc Information ware and software for systems}.
He was a  General Chair at the Moscow seminar
{\sc Information ware and software for systems} at the
Moscow House for Scientific and Technical Knowledge
Propagation (MHSTKP)

He was not only the principal scientific leader but also a generator
of the scientific ideas, moderator of the discussions and demonstrations
of the implemented toolkits and applications. His tutorial was listened by
some hundreds of humans, they occupied not only the seats in the Big Red
Conference Hall, but all the adjacent rooms and halls where
the radio translation was turned on.

During the tutorial duration there were no noise at all,
this was an actual, highly accented interest to the topic.
This was the best school of a live scientific investigation,
later many of the beginners and Ph.D. students were influenced
to work in this area.
Kuzin analyzed the advances and perspectives in
an area of the information systems which could be
treated as the Artificial Intelligence Systems (AIS).
He established the criteria and classification which
are applicable to characterize AIS among the manifold
of the Information Systems (IS), he analyzed the nature
of difficulties attempting the practical Knowledge Representation.
He did his tutorial in the engineering style, or using his
own terminology, engineering mathematician, entirely in
a constructive and even practical way, avoiding the troublesome
details of the ontology.

According his tutorial, the future of a matter was extremely promising
and scientifically interesting, and possible results of a work were
significant. And all of this were in detailed and rigorous correspondence
with the known facts and trends. He gave the proper engineering
approximation of a rather complicated idea for the properties
of a Knowledge Representation Language.
He demonstrated  his approach
to develop the software and to evolve the Software Engineering,
he pointed out what are
the promising challenges in the world known research projects and
what are the palliatives and followers of the ad hoc trends in development
and fundamental research.

Kuzin highly estimated an elegance in the mathematical results,
but did not like the unlimited usage of this elegance principle.
Often he accepted and estimated the results of implementation,
even without a rigorous mathematical background,
which was based on the sound intuitive ground. But this was just a first stage,
after which he, as a rule, inspired the mathematical verification
up to establishing the computational model.

\vspace{1ex}
\centerline{\bf 1979a}
\vspace{1ex}

\noindent Kuzin published, under his edition, the book of selected papers
in AI-field reflecting the state-of-art in USSR.
The title: {\sc Topics for Cybernetics. Intelligent Data Banks}, vol.55,
the USSR Academy of Sciences, Moscow, 1979.
This was an extremely popular book which was mentioned
among the specialists just as ``TC-55'' during half a decade.
Kuzin selected out, reviewed and edited the papers, with
an editing board, during almost half an year making his duties with a care.

This book gave insight and guidelines for both a theory and practice
of Intelligent Data Banks and other advanced Information Systems
which borrow the AI ideas. Kuzin formulated in depth the concept
and notion of IDB which later were used and enriched by the
hundreds of researches during the years. He accented that IDB is
a kind of Cybernetics System which includes the range of bases:
target base, knowledge base, data base, and, possibly some additional
bases.

Kuzin concluded that the Cybernetics Systems of this kind need
their own, newly established, mathematical foundations
which, possibly, have no immediate predecessors and violate
the traditional, e.g., continuous mathematics.
He was interested in establishing a measure or unit of knowledge
which give a sound basis for practical representation. He had a durable
interest to the frames' research area which gave a ground to the frameworks
based on scripts and situations selected out in the problem domains.
One of his ideas was to iterate some simple structure to obtain
the actual framework. He proposed to use the models for non-classical logics,
but, possibly, had not yet enough intuition how to overcome the contradictions
in the large bulk of information. In AI-field the research activity in this
direction just started, the difficulties were hardly understood by the specialists,
but no accepted constructive theoretical framework was established.
Due to John McCarthy et. al the systems of non-monotonic reasoning
with varying assumptions were established, so that some possibility to
eliminate the contradictions did exist. But Kuzin preferred to find out
some special algorithmic procedures or algebraic ground to re-solve
and avoid the contradictions towards consistency of the represented information,
instead of using and exploring the deductive problem solvers.

His principle paper on IDB inspired a great interest in Russia,
its importance is acknowledged till the current years.
It was conceptual in depth, enforcing the researchers to think over
the range, scale and scope of the possible theory for Intelligent Systems,
as well as over the limitations of the traditional Cybernetics and mathematical
models, where they cancel out their working scope and the modern
trends in cybernetics and Information Systems would affect.

Retrospectively speaking, many the researchers of 70s in AI-field
took their own roads and tracks of investigations, which previously
were hardly known, their theories met the scepticism from the orthodox
mathematicians. The discussions, initiated at the conferences,
had a lot of continuations in the voluminous publications of
the academic periodicals.

Some of the TC-55 papers were invited, and Kuzin personally
gave a lot of comments to the authors, improving the contents.
He stressed, that each of these papers must contain the elements
of foundations and guides to the future investigations. After the years passed,
the Kuzin's strategical plan was succeeded, attracting both a lot
of single volunteers and scientific groups. Almost all the young authors
of the papers became the known specialists, in some cases founding
their own scientific schools and directions of the research and development.
Their origin have an easy recovering up to IDB book.

\vspace{1ex}
\centerline{\bf 1979b}
\vspace{1ex}

\noindent {\sc The USSR Scientific and Technical Symposium on
Dialog and Factographic Systems for Information Services}.
This event was on October, 1979 in Ruza. Kuzin was invited by academician Germogen
Pospelov from the USSR Academy of Sciences  to organize the section. According the
agreement, he got the duties of the principal coordinator for the main
Artificial Intelligence
stream,   Knowledge Representation and Intelligent Data Banks. One of the main targets was to
start up in the USSR the practical development and applications of the intelligent information
systems, hence for Kuzin the organizing duties were dominant. And he succeeded in his
efforts.

An atmosphere of this Symposium was stimulating, the participants could enjoy
the golden autumn walking across the nearest forests.
All the sessions started at 10:00 AM but Kuzin was in the habit  to awake at 7:00 AM, so that all
his assistants did the same.
The working day of Kuzin's team started with a light running along
the bank of the forest river,
then sitting at the bench, they discussed the main topics of the current day
session, brief outline of the reports and possibilities to organize the evening panels.
It was a cool morning, the thoughts and proposals were generated distinctly,
becoming in the years the valid directions of a scientific research.

Some of the evening panels were moderated while most of them were the natural continuations
of the daytime sessions. The participants were attracted by Kuzin, his vision of the AI trends
when he estimated and analyzed the topics. In the late evenings, when the participants were tired,
he moderated the Russian
romance concerts singing solo ``Shine, shine, my star'', ``I met you'', and a lot
of others. His strong and deep voice attracted the additional participants and soon the hotel hall
became overfull.  The singing was accompanied by the piano music, the other singers changed
Kuzin or just the people listened to the classical music.

All of this was happened naturally, without preliminary planning, all the participants enjoyed the
intelligent freedom. But in a time the concert again was changed by the multiple local
discussions on how to build the Knowledge Systems.
Kuzin at ease directed all the discussions simultaneously, sometimes changing the topic or taking
his own initiative. These schools of contacts was stimulating for the researchers, especially for
Ph.D. students.

With a short time the people created the research teams, exchanging the papers, reports and
manuscripts between the teams. They were proud of belonging to Kuzin's school.

Kuzin's tutorial at the session gathered a  lot of participants, the conference hall was full.
He classified the numerous works and publications, using the qualitative evaluations, the
characteristics of efficiency and throughput, generalized and concluded, and estimated the
possible benefits and trends. He was confident that
the Intelligent CAD systems should contain,
besides the database and knowledge base, as a focal subsystem the target base. The most
important was to evolve its structure and features, and to solve the most complicated task   to
establish, implement and release the mathematical model.
His tutorial stimulated the research in the developing the
target bases which was done by many research teams.

At the panel discussions some speakers pointed out an interest for the knowledge representation
units, the frames, comparing the authors and their approaches. There were a lot of criticism and
sceptic attitude. Kuzin did not share this position,
it seemed that he has an intuitive vision of the
in-the-years perspective and long term trends. He usually was right when selecting out what to
investigate, what has a perspective and what lost the rate and is a blind alley.
In his research predictions he practically never was failed, this was known for other researchers,
and as a consequence his Ph.D. students had a good orientation and a starting position to succeed
in their own research. At this stage he even enforced the Ph.D. students to develop and
implement the knowledge systems based on the distinct notion of frame. He highly estimated the
results of a good experiment work which was compared with the same of the other authors. His
strategy was to accumulate a range of working prototypes of the intelligent information systems
with the aim to develop a full scale release for the industrial applications.

The same year (1979) he published his famous 2nd volume of
``Foundations of Cybernetics'' with a special subtitle
``Models for Cybernetics''. This book was written with a great
inspiration and confidence in a good perspective for the
Artificial Intelligence models during at least 10 years period.

In his book the most important and prominent models for knowledge based systems
were selected out and classified. The readership level was
as usually the engineering mathematician students, graduate
and undergraduate. He applied special efforts to avoid
the complicated formal derivations and used instead the
trasparent intuitive reasons, the feasibility and qualitative
measures, hints and explanations. He described and estimated the scopes
of the models and the possible modifications to capture
more meaning under the renewed conditions.

His approach attracted a lot of successors and inspired
a series of scientific discussions.

\vspace{1ex}
\centerline{\bf 1980}
\vspace{1ex}

\noindent {\sc The 2nd USSR Workshop on Intelligent Data Banks}.
This Workshop was on March~10--13, 1980 in Tsakhadzor, Armenia.
He was General Co{-}Chair sharing his duties with Dr. Ashot Hovanesyan,
a director of Medical Information and Computing Center of Armenia.
The efforts of joint organizing team to prepare this event, disseminate call for papers,
book the conference hall took several months.
Kuzin's position helped the organizers to overcome all the difficulties
and involve a lot of scientists
from various research institutions of the country.

This event was hosted by the team from
Medical Information and Computing Center of Armenia.
This was due to the attractiveness of the topics and the live interest
of the specialists in Medical Cybernetics to develop and apply the
intelligent information systems for the health care.
The computer-aided working place for the physician was one
of the nearest targets. The specialists in medical cybernetics were
granted by the government to make a progress in this area.
Most of them were highly interested to achieve the desired results,
there was a stimulating and collaborative atmosphere.

The participants were located in the hotel at a nice mountain resort,
they were vivid and  highly collaborative. Among them were
the known specialists with invited reports, researchers with different skill,
Ph.D. students, specialists from the industry.

Kuzin always was in a center of events, nobody had a free time.
The working day started early in the morning with the mountain skies
and sliding down the nearest hills. The participants used this opportunity
to establish the personal contacts which later generated the good
research teams.

The regular sessions were started after lunch time and
had a duration till the late evenings
while the participants just were tired to continue the panels.
A few late evenings were used for the on-fly concerts which were
in a Grand Red Hall. The participants could enjoy the classical piano music
and Armenian songs of the highly musically talented hosts.
Kuzin changed them
with the Russian romances.

The topic of his tutorial and moderated panels dealt with
the augmentation of the concepts for an information system
based on the Intelligent Data Bank. The ways of implementing
the adequate data models were at a central position.
He gave the feature analysis of the IDB projects and prototypes.
Some of the implementations were based on the dialects of
Frame Representation Language (FRL),
in a few projects the ideas of Nick Roussoupolos
to separate frames and relational database were used,
others were the followers of deductive models.
The team from Medical Information and Computing Center of Armenia
implemented and applied to the practical tasks the IDB
based on IMS, this was a balanced and successful implementation,
in their reports the results of two years experience
were discussed.

This was a highly stimulated event with the experience exchange,
the relations between many of the institutions were established.
This allowed later to share the job of collaborators for
the project ``Data Bank''.

\vspace{1ex}
\centerline{\bf 1981}
\vspace{1ex}

\noindent {\sc The 2nd USSR Scientific and Technical Symposium on
Dialog and Factographic Systems for Information Ware}.
This Symposium was held on April, 1981 in Suzdal,
its aims and scopes were analogous to the 1st
Symposium in 1979. Kuzin was a Chair of the Section on Knowledge Representation in
Intelligent Systems. He decided to attract the contributions in an area of CAD, and especially in
Advanced Programming Technology (APT). His aims were to discuss and analyze in depth the
Intelligent Programming Systems and Intelligent CAD. More briefly, he concentrated the efforts
on the target description generation for the system under development. His efforts at the 1st
Symposium helped him to attract to this event the high quality submissions. Many of the
researchers, under his influence, started to develop their frame-based software, dropping  the
initial system description down to the elementary units of knowledge representation.
His school now had the branches with the research teams from
Armenia, Baltic States, Georgia, the Ukraine.
The new results were reported by the team from Armenia,
they launched in the USSR the
applied Intelligent Information Systems for Medical Cybernetics. Later the joint project was
launched by the teams from Medical Information and Computing Center (MICC) of Armenia and
from MICC of the Russian Federation.

\vspace{1ex}
\centerline{\bf 1982}
\vspace{1ex}

\noindent {\sc The 3rd Workshop on Intelligent Data Banks}.
This workshop was in  Georgia.
He was a General Co{-}Chair sharing his duties with Profs. G.Chogowadze and G.Gogichaishwili
from the Georgian Politechnical Institute in Tbilisi, Georgia.
This was a joint forum where the
Georgian school in Artificial Intelligence was widely presented. The topics were as follows:
knowledge bases and their development, augmenting the relational databases with the intelligent
access, interfaces, intelligent CAD, and various applications. Most of the Georgian research
teams shared Kuzin's approach and worked in a close collaboration with Kuzin's school.

In his tutorial Kuzin paid attention to the computational environment of IDB and its properties.
He proposed to organize on-fly the series of panel discussions and a round table to better
coordinate the mutual research activity.
He felt a good reason and opportunity to launch the
concert projects which were aimed towards development the
full scale prototypes of IDB. Some of
them must use the various data models and knowledge representation schemata, while others
used the specific problem orientation. The agreements between the institutions were achieved
and in the nearest future the projects were launched. This was a great advance in the sphere of
Artificial Intelligence in the USSR which strongly influenced the research activity in the country.

\vspace{1ex}
\centerline{\bf 1983}
\vspace{1ex}

\noindent {\sc The Coordinating Workshop of the Projects ``Data Bank'' and ``Situation'' on
Formal Models for Developing the Databases and Data Banks}.
This workshop was on October 25-27, 1983 in Uzhgorod. Kuzin was a Co{-}Chair as the principal
investigator for the project ``Data Bank''. This event was addressed to further stimulating of the
younger generation researchers. The reports were concentrated on the discussions and feature
analysis and trends for developing the advanced data models.

\vspace{1ex}
\centerline{\bf 1984a}
\vspace{1ex}

\noindent {\sc The 4th Workshop on Intelligent Data Banks}.
He was a General Co{-}Chair with Georgian partners the same way as at 3rd Workshop.
This Workshop was in Terscol, Naltchik and was planned in 1982. Its aim was to estimate the
results of joint projects in an area of IDB.

In his tutorial he analyzed the most advanced results of the research teams pointing out the
success in applying the relational databases. The Georgian teams accumulated a good experience
developing the intelligent applications for the industry. One of the projects from the Russian team
resulted in a toolkit as the extensible computation environment with the embedded knowledge
base system and relational database system. The basic primitives were objects within the host
programming system LISP, implemented by the system programmer Anatoly Panteleev. The
features of implementation and applications were discussed at the evening panels.

Among the interesting theoretical issues on data models were the categorical model and tensor
model. The research teams which worked in this directions believed in the promising results and
demonstrated the embedding of the relational algebra into categorical or tensor framework. They
were at the beginning stage with the experiments, only partial implementations were released.

As was happened, one of the main topics for this Workshop were the embedded systems.

\vspace{1ex}
\centerline{\bf 1984b}
\vspace{1ex}

\noindent {\sc The International Conference on
Databases in Networks}.
This event was on February, 1984 in Moscow, hosted by the International Center for Scientific
and Technical Information. One of the main topics was the AI methods for the
New Information
Technologies.
He was an invited speaker and in his tutorial gave in depth analysis of the most
prominent results in AI field and their fruitfulness for the Advanced Information Technologies.

\vspace{1ex}
\centerline{\bf 1986}
\vspace{1ex}

\noindent {\sc The 1st Workshop on Intelligent Access for Data Banks}.
This workshop on November~19-21, 1986 was in Dilijan, Armenia,
with Kuzin as a General
Co-Chair.
Dr. Ashot Hovanesyan was both a General Co{-}Chair and Program Committee Chair and
attracted a set of submissions reflecting the state-of-art in an area. Kuzin's school was presented
by the numerous reports. All the relative research teams also joined this event. In fact, this was
the 5th Joint Workshop on Intelligent Data Banks which continued the series of previous four
workshops.

Kuzin's tutorial ``Evolution, advances and trends in an area of intelligent access for data banks''
later was disseminated in manuscript within the research community. One of his ideas was to use
objects as data and knowledge units which are equipped with the attached procedures. Under this
assumption the information system would be extensible. This time he worked at the
extensible
data models and gave their feature analysis in depth.

 \section*{Acknowledgements}

I would like to thank all the colleagues and collaborators
for valuable help and advice during the preparation of this paper.


\begin{thebibliography}{4}
\bibitem{ref1}Wolfengagen~V.E. L.T.~Kuzin: Research Program.
JurInfoR-MSU Electronic Library,
{\tt http://msu.jurinfor.ru/e-lib/}
\end{thebibliography}



\end{document}